\documentclass{appolb}
\usepackage{epsfig}

\newcommand{\hep}[1]{{\tt hep-ph/#1}}

\newcommand\jp[3]{{ J.~Phys. }{\textbf #1} (#2) #3}
\newcommand\npb[3]{{ Nucl. Phys. }{\textbf B #1} (#2) #3}
\newcommand\npbps[3]{{ Nucl. Phys. }{\textbf B} { (Proc. Suppl.)}{ \textbf #1} (#2) #3}
\newcommand\plb[3]{{ Phys. Lett. }{\textbf B #1} (#2) #3}                   
\newcommand\prd[3]{{ Phys. Rev. }{\textbf D #1} (#2) #3}            
\newcommand\zpc[3]{{ Z. Physik }{\textbf C #1} (#2) #3}             
\newcommand\sjnp[3]{{ Sov. J. Nucl. Phys. }{\textbf #1} (#2) #3}    
\newcommand\jetp[3]{{ Sov. Phys. JETP }{\textbf #1} (#2) #3}        
 
\newcommand{\as}{\alpha_{\mathrm{s}}}

\newcommand{\dJ}[1]{\frac{\dif\sigma}{\dif J}{}\raisebox{1.5ex}{$#1$}}
\newcommand{\dif}{{\rm d}}

\newcommand{\dug}{\,\raisebox{0.37pt}{:}\hspace{-3.2pt}=}
\newcommand{\esp}[1]{{\rm e}^{#1}}
\newcommand{\fz}{f^{(0)}}
\newcommand{\hz}{h^{(0)}}
\newcommand{\id}{\mbf{1}}

\newcommand{\kd}{\kk_2}
\newcommand{\kk}{\mbf{k}}
\newcommand{\ku}{\kk_1}

\newcommand{\mbf}[1]{\mbox{\boldmath $#1$}}

\newcommand{\pa}{{\mathsf a}}
\newcommand{\pb}{{\mathsf b}}
\newcommand{\pg}{{\mathsf g}}

\newcommand{\pq}{{\mathsf q}}

\newcommand{\qq}{\mbf{q}}

\newcommand{\ui}{{\mathrm i}}

\newcommand{\lab}[1]{\label{#1}}
\newcommand{\labe}[1]{\label{#1}}
\begin{document}
\title{Jets at high energies, factorization \\ and jet vertex in NL ln(s)
\thanks{X International Workshop on Deep Inelastic Scattering DIS2002, \\
Cracow 30 April - 4 May 2002 }%
}
\author{G. P. Vacca
\address{Dipartimento di Fisica, Universit\`a di Bologna and
Istituto Nazionale di Fisica Nucleare, Sezione di Bologna,
via Irnerio 46, 40126 Bologna, Italy}
}
\maketitle
\begin{abstract}
The next-to-leading corrections to the jet vertex which is relevant for
the Mueller-Navelet jets production in hadronic collisions and for the
forward jet cross section in lepton-hadron collisions are presented in the
context of a $k_t$ factorizazion formula which resums the leading and
next-to-leading logarithms of the energy.
Both the quark- and gluon-initiated contribution are now computed.
This completes the framework for a full phenomenological analysis of
Mueller-Navelet jets in NL log(s) approximation. Forward jets
phenomenology still requires the NL photon impact factor.
\end{abstract}
\section{Introduction}
In recent works \cite{BaCoVa01,BaCoVa02} a novel element, relevant in the
study of QCD in the Regge limit, has been defined and computed at the
NLO level. It is the jet vertex, which represents one of the building
blocks in the production of Mueller-Navelet jets~\cite{MuNa87} at hadron
hadron  colliders and of  forward jets~\cite{Mu90} in deep inelastic electron
proton scattering. Such processes should provide a kinematical environments
for which the BFKL Pomeron~\cite{BFKL76} QCD analysis could apply, since the
transverse energy of the jet fixes a perturbative scale and the large energy
yields a large rapidity interval.

We briefly remember here that in a strong
Regge regime important contributions, or even dominant, come, in the
perturbative language, from diagrams beyond NLO and NNLO at fixed order in 
$\alpha_s$. This is the main reason for considering a resummation of the
leading and next-to-leading logarithmic contributions as
computed in the BFKL Pomeron framework. Such approach is lacking of
unitarity so that, if the related corrections are not taken into account,
one must consider an upper bound on the energy to suppress them.
It is already known that the LL analysis is not accurate
enough~\cite{nonasympt}, being the
kinematics selected by experimental cuts far from any asymptotic regime.
Moreover at this level of accuracy there is a maximal dependence in the
different scales involved (renormalization, collinear factorization
and energy scales). For the Mueller-Navelet jet production process the only
element still not known at the NLO level was the ``impact factor'', which
describes the hadron emitting one inclusive jet when interacting with the
reggeized gluon which belongs to the BFKL ladder, accurate up to
NLL~\cite{FaLi98,CaCi98}. The jet vertex, now computed,
is the building block of this interaction.
For the so called forward jet production in DIS the extra ingredient
necessary is the photon impact factor, whose calculation is currently
in progress~\cite{BaGiQi00, FaMa99}.  
Let us also remind that NLL BFKL approach has recently
gained more theoretical solidity since the bootstrap condition in its strong
form, which is the one necessary for the self-consistency of the
assumption of Reggeized form of the production
amplitudes, has been stated~\cite{bootstrapnloBV} and
formally proved~\cite{FaPa02}. This relation is a very remarkable
property of QCD in the high energy limit.

\begin{figure}[hb!]
\centering
\resizebox{0.3\textwidth}{!}{\includegraphics{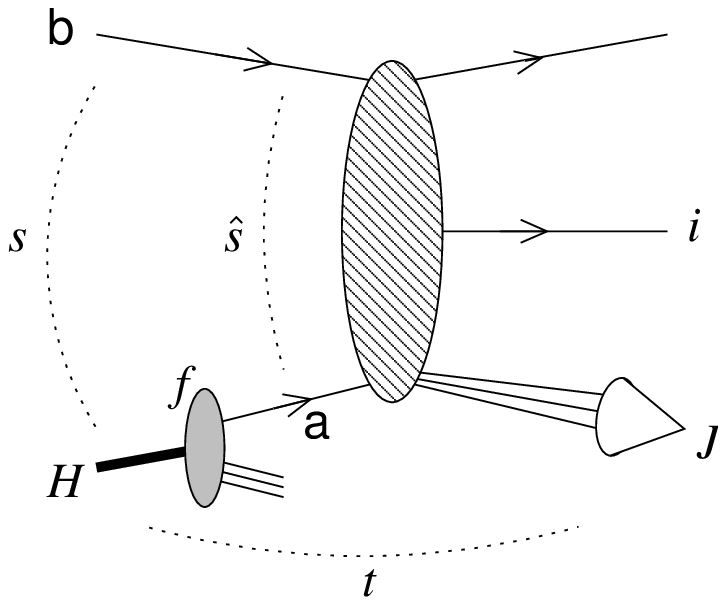}}
\caption{\labe{f:jet} High energy process with jet production.
$H$ is the incoming hadron providing a parton $\pa$ (gluon $\pg$/quark $\pq$)
with distribution density $f_\pa$ which scatters with the parton $\pb$
$J$ denotes the jet produced in the forward direction (w.r.t $H$) and
$i$ is the generic label for outgoing particles.}
\end{figure}

A particular theoretical challenge, interesting by itself and
appearing in the calculation, is related to the special kinematics.
The processes to be analyzed is illustrated in Fig.\ref{f:jet}: the lower parton
emitted from the hadron $H$ scatters with the upper parton
$\pq$ and produces the jet $J$.
The gluon is hard, because of the large transverse momentum of the jet, 
and obeys the collinear factorization, i.e.
its scale dependence is described by the DGLAP evolution equations \cite{DGLAP77}.
Above the jet, on the other hand, the kinematics chosen requires a large rapidity
gap between  the jet and the outgoing parton $\pq$: such a situation is described
by BFKL dynamics. Therefore the jet vertex lies at the interface between
DGLAP and BFKL dynamics, a situation which appears for the first time in
a non trivial way. As an essential result of our analysis we find 
that it is possible to separate, inside the jet vertex, the collinear infrared 
divergences that go into the parton evolution of the incoming gluon/quark from 
the high energy gluon radiation inside the rapidity gap which belongs to 
first rung of the LO BFKL ladder.

\section{Jet vertex and cross sections}
Let us consider the kinematic variables
$p_H = \left(\sqrt{s/2},0,\mbf{0}\right)$, $s\dug(p_H+p_\pq)^2$,
$p_\pb = \left(0,\sqrt{s/2},\mbf{0}\right)$,
$p_i =E_i\left(\esp{y_i}/\sqrt2,\esp{-y_i}/\sqrt2,\mbf{\phi}_i\right)$,
$p_\pa = x\,p_H$.
In our analysis we study the partonic subprocess $\pa+\pb\to X+{\it
jet}$ in the high energy limit
\begin{equation}\lab{HElimit}
 \Lambda_{\rm QCD}^2 \ll E_J^2\sim -t\; {\rm (fixed)} \ll s\to\infty
\end{equation}
According to the parton model, we assume the physical cross section to be given by the
corresponding partonic cross section $\dif\hat\sigma$ (computable in perturbation theory)
convoluted with the parton distribution densities (PDF) $f_\pa$ of the
partons $\pa$ inside
the hadron $H$. A jet distribution $S_J$, with the usual safe infrared
behaviour, selects the final states contributing to the one
jet inclusive cross section that we are considering.
In terms of the jet variables --- rapidity, transverse energy and azimuthal
angle ---  the one jet inclusive cross section initiated by quarks and
gluons in hadron $H$ can be written as
\begin{equation}\lab{pSf}
\dJ{} \dug \frac{\dif\sigma_{\pb H}}{\dif y_J \dif E_J \dif \phi_J}
= \sum_{\pa=\pq,\pg} \int\dif x\;
\dif\hat\sigma_{\pb\pa}(x)\,S_J(x)\fz_\pa(x)\;.
\end{equation}

One can easily see~\cite{BaCoVa01,BaCoVa02} that at the {\it lowest order}
the jet cross section, dominated by a $t$-channel gluon exchange, can be written as
\begin{equation}\lab{LOFF}
 \dJ{^{(0)}} = \sum_{\pa=\pq,\pg}
\int\dif x \int \dif\kk\;\hz_\pb(\kk)V^{(0)}_\pa(\kk,x)\fz_\pa(x)
\end{equation}
where $V^{(0)}_\pa(\kk,x)=\hz_\pa(\kk)
S_J^{(2)}(\kk,x)$ is the jet vertex induced
by parton $\pa$, $\hz_\pa(\kk)$ is the partonic impact
factor and $\fz_\pa(x)$ is the parton distribution density (PDF).
The jet distribution is in this case trivial,
$S_J^{(2)}(\kk,x)= \delta\left(1-x_J/x
\right)E_J^{1+2\epsilon} \delta(\kk-\kk_J)$ with
$ x_J \dug E_J\esp{y_J}/\sqrt s$.

At the {\it NLO approximation} virtual and real corrections enter in the
calculation of the partonic cross section
$\dif\hat\sigma_{\pb\pa}$. The three partons produced in the real
contributions, in the upper, and lower rapidity region are denoted by
$2$ and $1$, while the third, which can be emitted everywhere, by
$3$. Moreover we shall call $k=p_\pb-p_2$ and $k'=p_1-p_\pa$ and
$q=k-k'$. Bold letters as before indicate the trasverse part.
The infrared and ultraviolet divergences
can be, as usual, treated by dimensional regularization ($d=4+2\epsilon$).
Taking into account the I.R. properties of the jet distribution
$S_J^{(3)}$  the following structure is matched exactly up to NLO
(i.e. $\alpha_s^3$)~\cite{BaCoVa01,BaCoVa02}
\begin{equation}
\dJ{}= \sum_{\pa=\pq,\pg} \int\dif x \int \dif\kk\,\dif\kk'\;
 h_\pb(\kk) G(xs,\kk,\kk') V_\pa(\kk',x) f_\pa(x)
\label{jetcrosssection}
\end{equation}
where
$h = \hz + \as h^{(1)}+\cdots$, $V= V^{(0)} + \as V^{(1)}+\cdots$,
$f = \fz + \as f^{(1)}+\cdots$ and
$G(xs,\kk,\kk') \dug \delta(\kk-\kk')+\as
K^{(0)}(\kk,\kk')\log\frac{xs}{s_0}+\cdots$.
The partonic impact factor correction in forward direction $h^{(1)}$
is well known~\cite{CiCo98}, the PDF's $f_\pa$ are the standard ones
satisfying the LO DGLAP evolution equations and the BFKL Green function
$G$ is defined by the LO BFKL kernel $K^{(0)}$. The new element
is the correction to the jet vertex $V^{(1)}$ whose
expression, for the quark and gluon initiated case, is given in~\cite{BaCoVa01,BaCoVa02}. 

Another element, crucial in the derivation of the
representation given above, is the energy scale $s_0$
associated to the BFKL rapidity evolution. The calculations show a
natural choice, due to angular ordered preferred gluon emission and
the presence of the jet defining distribution, which
is also crucial to obtain the full collinear singularities which
factorize into the PDF's. We give the expression of the jet vertex for
the case $s_0(\kk,\kk') \dug (|\kk'|+|\qq|)(|\kk|+|\qq|)$.
A mild modification of such a scale can be
performed without introducing extra singularities, but in general this
is not true. In any case using a different scale requires the
introduction of modifing terms. For the
symmetric Regge type energy scale $s_R=|\kk||\kk'|$, one has
\begin{equation}
 G(xs,\kk,\kk')=(\id+\as H_L)\left[\id+\as K^{(0)}\log\frac{xs}{|\kk||\kk'|}\right]
(\id+\as H_R),
\end{equation}
where
$H_L(\kk,\kk')=-K^{(0)}(\kk,\kk')\log\left(|\kk|+|\qq|)/|\kk|\right)=H_R(\kk',\kk)$.

To obtain the jet cross section with accuracy up to NLL terms, one has
to consider the NLL BFKL kernel $K = \as K^{(0)}+\as^2 K^{(1)}$, which has been computed
with the scale $s_R=|\kk||\kk'|$. The corresponding Green function, to be used
in (\ref{jetcrosssection}), is given by
\begin{equation}
G(xs,\ku,\kd) =
\int\frac{\dif\omega}{2\pi\ui}\left(\frac{xs}{s_R}\right)^\omega
\langle\ku|(\id+\as H_L) [\omega - K]^{-1}(\id+\as H_R)|\kd\rangle\,.
\end{equation}
The formula for the Mueller-Navelet jets~\cite{BaCoVa02} can be easily
derived symmetrizing the formula (\ref{jetcrosssection}) for the two
jet case.
\section{Conclusions}
The calculations at the NLO accuracy for the jet vertex which lies at
the interface between DGLAP and BFKL dynamics are completed. Therefore
a phenomenological analysis of the Mueller-Navelet jets at NLL is now possible.
\section*{Acknowledgments}
The results discussed have been obtained in collaboration with J. Bartels
and D. Colferai~\cite{BaCoVa01,BaCoVa02}.

\end{document}